\documentclass[aip, graphicx, reprint,superscriptaddress,amsmath,amssymb]{revtex4-1}
\usepackage{color}
\usepackage{textcomp}
\usepackage{graphicx}
\usepackage[colorlinks=true, urlcolor=blue, linkcolor=blue, citecolor=blue]{hyperref}

\begin{document}
\title{Two-dimension Dirac fermions system in CdHgTe quantum wells} 

\author{M.\,L.\,Savchenko}
 \email{SavchenkoMaximL@gmail.com}
\author{D.\,A.\,Kozlov}
\author{Z.\,D.\,Kvon}
\author{N.\,N.\,Mikhailov}
\affiliation{Novosibirsk State University, Novosibirsk 630090,
Russia} \affiliation{Rzhanov Institute of Semiconductor Physics,
Novosibirsk 630090, Russia}

\author{S.\,A.\,Dvoretsky}
\affiliation{Rzhanov Institute of Semiconductor Physics,
Novosibirsk 630090, Russia}

\author{B.\,A.\,Piot}
\affiliation{Laboratoire National des Champs Magn\'etiques
Intenses, Grenoble 38042, France}

\date{\today}

\begin{abstract}
We report on transport and capacitance spectroscopy study of two
kinds of quantum wells, namely Cd$_{0.02}$Hg$_{0.98}$Te and
Cd$_{0.06}$Hg$_{0.94}$Te with the thicknesses of 7.4 and 11.5\,nm,
accordingly. The fraction of Cd was chosen in a way that the both
quantum wells are expected to have gapless band structure typical
for a Dirac fermions system. We have established that the first
quantum well exhibits a massless Dirac fermions system with a
quality slightly better then in conventional HgTe quantum wells of
critical thickness. Second quantum well exhibits a high-quality
two-dimensional topological insulator state with the energy gap of
around 10\,meV and well-defined edge transport making it as a good
candidate for further study and applications of topological
insulators.
\end{abstract}

\maketitle 

\section*{Introduction}
The two-dimensional single valley gapless Dirac Fermions (DF)
system implements in a HgTe quantum well (QW) with a critical
thickness $d$ of around 6.5\,nm \cite{Raichev2012}. The band
diagram of the system is very similar to graphene but without the
valley degeneracy and with Dirac point located in a center of the
Brillouin zone. The system consists of a great interest for a
condensed matter physics because of the wide variety of effects
demonstrated. It was studied by transport measurements both in
classical and quantizing magnetic fields \cite{Buttner2011,
Tkachov2011a, Kozlov2012, Dobretsova2016a, Kozlov2014a,
Kozlov2014b}, the linearity of it's band structure was shown by
cyclotron resonance studies in THz range \cite{Kvon2011,
Olbrich2013a,Zoth2014} as well as by the capacitance spectroscopy
\cite{Kozlov2016c}. The striking results specific for DF system
were obtained by Faraday rotation study, where the quantization of
the rotation angle in the units of fine structure constant was
observed \cite{Shuvaev2016} and by non-local transport
measurements \cite{Gusev2017}, where the existence of edge mode
with the filling factor of $i=0$ was shown.

Further study of the system is limited by several issues,
including the disorder which hinders the observation of subtle
effects and reduces the carriers' mobility. One could note the
following regularity of HgTe QWs: the electron mobility increases
with QW's thickness. Indeed, the electrons in QW of critical
thickness demonstrate a moderate mobility $\mu$ not exceeding
$1.5\cdot 10^5$\,cm$^2$/V$\cdot$s \cite{Dobretsova2016a} which
becomes even smaller for thinner QWs ($2.5\cdot10^4$ for $d=
5.7$\,nm \cite{Hubmann2018}), while QWs with $d>10$\,nm could
demonstrate the mobilities almost one order bigger
\cite{Tkachov2011a,Dobretsova2015a}. The possible reason for the
observed $\mu(d)$ dependence is the spatial fluctuations of the
QW's thickness $\Delta d$ leading to the energy gap
inhomogeneities and additional scattering
\cite{Tkachov2011a,Dobretsova2015a}. The thickness fluctuations is
expected to be independent from $d$, making thinner QWs more
subjected to that kind of imperfectness since the relative
fluctuations $\Delta d/d$ are bigger. While the issue itself
desires a separate study, the fact that thicker QWs shows higher
electron mobility could be used in order to increase the mobility
of DF system. On the one hand, the gapless state implements in QWs
of critical thickness $d_\text{c}$ corresponding to the transition
from normal band structure to the inverted one \cite{Raichev2012}.
The value of $d_\text{c}$ depends on the QW orientation and strain
but it is well-defined and lies in the range of 6.3-6.6\,nm.
However, the value of $d_\text{c}$ could be efficiently increased
if one replaces a part of Hg atoms in the HgTe QW with Cd ones. In
particular case of Cd$_{0.17}$Hg$_{0.83}$Te alloy the critical
thickness tends to the infinity and DF are formed in bulk material
\cite{Orlita2014, Teppe2016}.

In the current work we report on the magnetotransport and
capacitance spectroscopy study of two kinds Cd$_x$Hg$_{1-x}$Te QWs
with the thickness of $d = 7.4$\,nm (type 1 QW) and $d = 11.5$\,nm
(red{type 2 QW}). The fraction of Cd was chosen in a way that the
both QWs are expected to have gapless band structure typical for
DF system. However the found that type 2 QW is characterized by a
small energy gap and an inverted band structure typical for
two-dimensional topological insulators. Both QWs demonstrate
higher values of electron mobility, than conventional pure HgTe
QWs with the critical thickness of around 6.5\,nm.

\section*{Methods}
\begin{figure}
    \includegraphics[width=1\columnwidth]{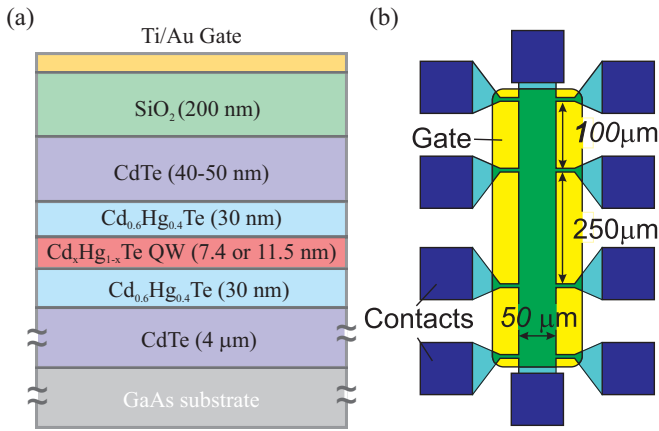}
    \caption{
        (a) - The schematic cross-section of the structures under study.
        The Cd$_{x}$Hg$_{1-x}$Te film (type 1 QW with thickness 7.4\,nm for
        $x=0.02$
        and type 2 QW with thickness 11.5\,nm for $x=0.06$)
        is sandwiched between thin
        Cd$_{0.6}$Hg$_{0.4}$Te buffer layers, covered by 40-50\,nm of CdTe
        and a 200\,nm SiO$_2$ insulator and a metallic gate.
        (b) - The schematic top view of studied Hall bars.
    } \label{figSamples}
\end{figure}

We study Cd$_{0.02}$Hg$_{0.98}$Te (HgTe with 2\% of Hg atoms
replaced by Cd, type 1 QW) and Cd$_{0.06}$Hg$_{0.94}$Te (6\% of
Cd, type 2 QW) quantum wells with the thicknesses of 7.4 and
11.5\,nm, accordingly. The structures have been grown by molecular
beam epitaxy on GaAs(013) substrate (Fig.~\ref{figSamples}~(a)).
Wet chemical etching was used to make 10-contacts Hall bars
(Fig.~\ref{figSamples}~(b)). The central part of the Hall bars was
covered by 200\,nm thick SiO$_2$ insulator and Ti/Au gate. Several
samples from the same wafers have been studied showing similar
results.

The measurements were performed in a temperature range of 1.8~--
50\,K and in perpendicular magnetic fields up to 3\,T. The
magnetotransport data were obtained using standard 4-terminal
lock-in technique with the excitation current within the range of
20 -- 200\,nA (depending on the sample resistance and temperature)
and the frequency of 12\,Hz. The capacitance was measured by
applying a sum of DC voltage $V_\text{g}$ and a small probing AC voltage
$V_\text{ac} = 20$\,mV at frequencies of 10 -- 700\,Hz to the gate and
measuring the AC current flowing through the QW. Both real and
imaginary parts of the signal were recorded in order to avoid
leakages and resistive effects.

\section*{Results and discussion}

We performed the magnetotransport study for both types of QW. The
analysis of the data obtained from each QW was pointed to the
further questions: 1. Does the QW have an energy gap? 2. What is
the maximum value of the electron mobility? Additionally we
performed the capacitance $C$ spectroscopy of the first type QW.
From the $C(V_\text{g})$ dependence we extracted the velocity of
DF and also estimated the magnitude of disorder at Dirac point.
For the second QW the capacitance spectroscopy was not performed
since we have discovered the presence of energy gap between
conduction and valence bands. In order to extract a value of the
gap and to check if QW is characterized by normal or inverted band
structure we additionally measured the resistance in non-local
geometry and checked the temperature dependence of the resistance
maximum both in local and non-local geometries. Next we present
the detailed analysis of the experimental data.

\subsection{Type 1 QW }

\begin{figure}
    \includegraphics[width=1\columnwidth]{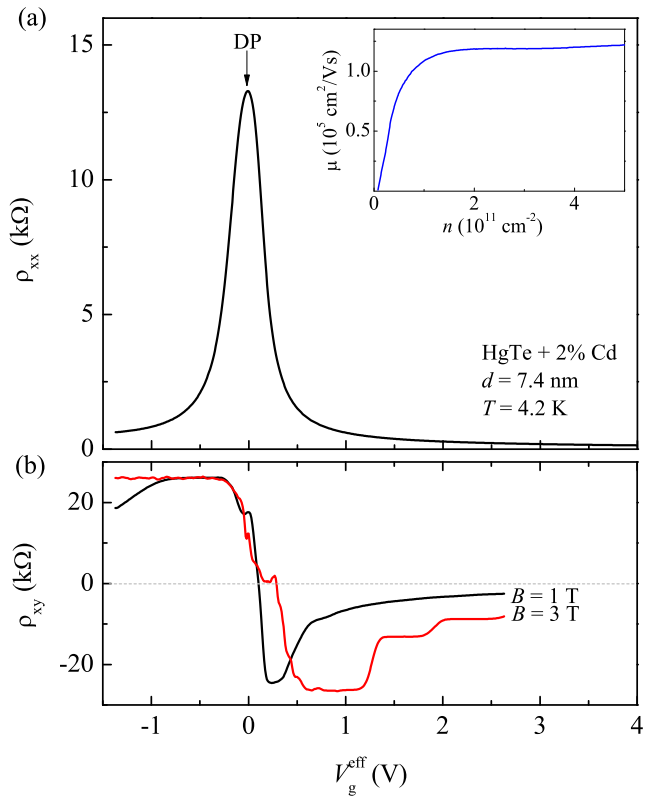}
    \caption{Type 1 QW.
        (a) - The effective gate voltage $V_\text{g}^\text{eff}$
        dependence of the sheet resistance
        $\rho_\text{xx}$ measured at $T = 4.2$~K and zero magnetic field.
        \textit{Inset} -- the density dependence of the electron mobility.
        (b) - The gate voltage dependence of the Hall resistance $\rho_\text{xy}$
        measured at the same temperature at two values of perpendicular
        magnetic field, 1 and 3\,T.
    }\label{fig:7.4nm_transport}
\end{figure}

Fig.~\ref{fig:7.4nm_transport}~(a) shows the gate dependence of
the sheet resistance $\rho_\text{xx} (V_\text{g}^\text{eff})$ of
7.4~nm HgTe quantum well with 2\% of cadmium measured at
$T=4.2$\,K, where $V_\text{g}^\text{eff} = V_\text{g} -
V_\text{g}^\text{DP}$ and $V_\text{g}^\text{DP}$ is the gate
voltage corresponding to the Dirac point. Note that the value of
$V_\text{g}^\text{DP}$ is determined by the random charge trapped
in the insulating layer and may vary from the cooling cycle to
cycle. However, every measured dependence is highly reproducible
if shifted with respect to $V_\text{g}^\text{DP}$ and plotted
versus $V_\text{g}^\text{eff}$. At $V_\text{g}^\text{eff} =0$ the
resistivity reaches its maximum with the value of about
13.3~k$\Omega \approx 0.5$\,$h/e^2$. The obtained value and
general dependence agree with the results from previous works on
conventional HgTe QWs of critical thickness \cite{Buttner2011,
Kozlov2014b, Gusev2017} and points out to the gapless band
structure. Fig.~\ref{fig:7.4nm_transport}~(b) shows the gate
dependence of the Hall resistance $\rho_\text{xy}$ in
perpendicular magnetic fields $B = 1$~T (black) and 3~T (red). In
the vicinity of DP the $\rho_\text{xy}(V_\text{g}^\text{eff})$
dependence changes its sign. On the right from the DP there is an
electron side, while hole side is on the left. From the classical
hall resistance, measured at smaller magnetic field (not shown),
we obtained linear dependence of density $n(V_\text{g})$. The
filling rate was found to be of about $9.2 \times
10^{10}$\,cm$^{-2}$/V. The inset in Fig.~\ref{fig:7.4nm_transport}
demonstrates $\mu(n)$ trace. Both shape of the dependence and the
mobility values are typical for DF in HgTe \cite{Dobretsova2016a}.

At strong magnetic field ($3$\,T) one could clearly recognize
several quantum Hall plateaux for electron filling factors $i$
from 1 to 3 and much longer hole plateux with corresponding $i =
-1$. The asymmetry between electron and holes plateau is
remarkable. Moreover, at 1\,T there are only hints of
$\rho_\text{xy}$ quantization on the electron side, whereas there
is clear $h/e^2$ plateau on the hole one. The similar effect was
observed in conventional HgTe QWs of critical
thickness\cite{Kozlov2014b} and was explained by the existence of
side valleys in the valence band, situated on some distance
$E_\text{HH}\approx 30$\,meV below the DP. When the Fermi level
touches the top of the side valleys an additional kind of heavy
holes emerges in the system. The heavy holes efficiently screen
the charge disorder and thereby increasing scattering time for
Dirac holes. At the same time the existence of heavy holes
strongly decreases the partial filling rate of Dirac holes and
prolongs the quantum Hall plateau for $i = -1$. One could suggest
that our system could be characterized by the similar band
structure with side valleys and therefore demonstrate the same
kind of behavior.

\begin{figure}
    \includegraphics[width=1\columnwidth]{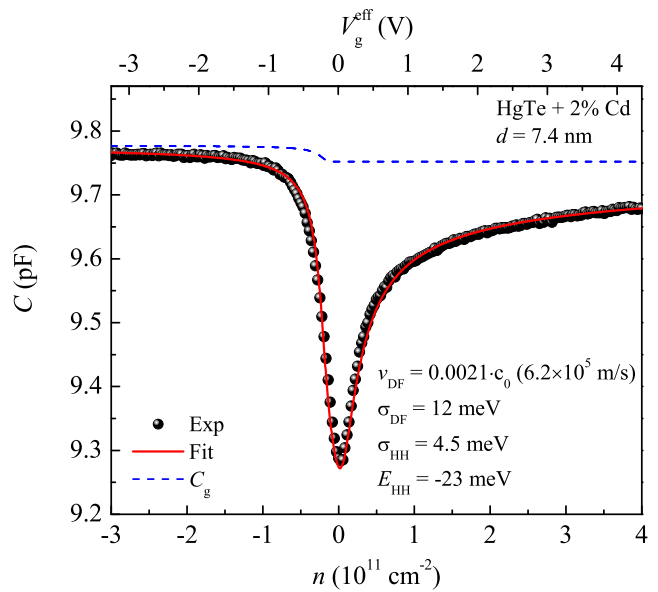}
    \caption{Type 1 QW.
        Experimental (black symbols) and theoretical (a red line) density
        dependence of the capacitance.
        The blue dashed line is the supposed geometric capacitance.
    } \label{fig:7.4nm_C}
\end{figure}

In order to quantitatively compare the value of disorder in the QW
under investigation and in conventional QWs we performed the
capacitance spectroscopy study. The capacitance measured between
the gate and QW could be represented as two capacitors connected
in series: geometric capacitance $C_\text{g}$ and quantum
capacitance $C_\text{q}=e^2D$, where $D$ is the theromdynamic
density of states (DoS) of the system\cite{Smith1985}.
Fig.~\ref{fig:7.4nm_C} shows the gate voltage (top axis) or
density (bottom axis) dependence of measured (black symbols) and
fitted capacitance (a red line). The variation of $C$ reflects the
change of DoS with density. In ideal DF system the DoS and
consequently $C$ should drop to zero in DP. In real case the spatial
fluctuations of trapped charge lead to the fluctuation of
carriers' density. The fluctuations does not allow DoS to reach
zero and modifies the shape of $C_\text{g}(n)$ trace in the
vicinity of DP\cite{Ponomarenko2010, Kozlov2016c}. We fitted the
obtained dependence with our model which takes into account both
the properties of the band diagram (linear dispersion law and the
existence of side valleys in the valence band) and disorder in the
system. The fitting model and procedure is explained in details in
Ref.~\onlinecite{Kozlov2016c}.

From the fitting one extracts the following parameters:
$v_\text{DF}$, the velocity of DF; $E_\text{HH}$, the distance
from DP to the top of the side valleys; $\sigma_\text{DF}$ and
$\sigma_\text{HH}$, the spatial fluctuations of the Fermi energy
in the vicinity of DP and side valleys accordingly. The values of
the parameters extracted from the fitting are shown in the
Fig.~\ref{fig:7.4nm_C}. The "band diagram parameters"
$v_\text{DF}$ and $E_\text{HH}$ are found to be slightly smaller
compare to previously reported values \cite{Olbrich2013a,
Kvon2011, Kozlov2016c}. What is essential that the "disorder
parameters" $\sigma_\text{DF}$ and $\sigma_\text{HH}$ are also
smaller in comparison to conventional HgTe QWs of critical
thickness ($\sigma_\text{DF}$ has a value of 16 / 12\,meV and
$\sigma_\text{HH}$ of 5.1 / 4.5~meV for the QW under investigation
/ conventional QWs, accordingly) \cite{Kozlov2016c}. That gives a
confirmation that the quality of HgTe QWs could be increased by
using thicker QW.

\subsection{Type 2 QW}

In our work we also investigated second type QW which consists of
Cd$_{0.06}$Hg$_{0.94}$Te alloy (HgTe with 6\% of Cd) with the
thickness of 11.5\,nm and performed the same transport
measurements as for the QW of type 1. The
$\rho_\text{xx}(V_\text{g}^\text{eff})$ dependence for the type 2
QW is shown in Fig.~\ref{fig:11.5nm_Rxx, Rxy}~(a). While its
general behavior is very similar to the previous case, the
resistivity value in the maximum for the type 2 QW is much bigger
(90\,k$\Omega$ $\approx$ 3.5\,$h/e^2$) indicating to the presence
of the energy gap between conduction and valence bands. The second
difference between QWs could be found in
$\rho_\text{xy}(V_\text{g}^\text{eff})$ dependence, shown in
Fig.~\ref{fig:11.5nm_Rxx, Rxy}~(b). In contrast to the first QW,
the quantized plateaux in $\rho_\text{xy}$ of the second QW on the
electron side becomes visible already at $B=1$\,T and completely
absent on the hole side. The detailed study of quantum Hall effect
in new QW is out of scope of the research.

\begin{figure}
    \includegraphics[width=1\columnwidth]{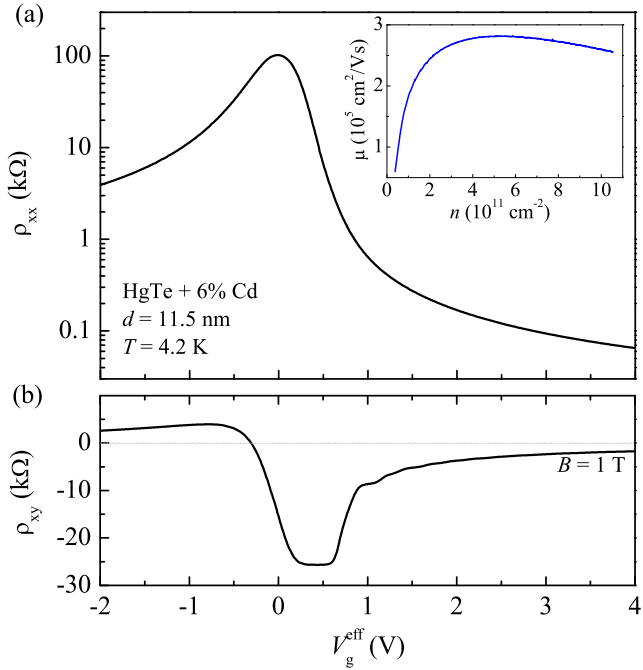}
    \caption{Type 2 QW.
        (a) - The effective gate voltage $V_\text{g}^\text{eff}$
        dependence of the sheet resistance
        $\rho_\text{xx}$ measured at $T = 4.2$~K and zero magnetic field.
        \textit{Inset} -- the density dependence of the electron mobility.
        (b) - The gate voltage dependence of the Hall resistance $\rho_\text{xy}$
        measured at the same temperature and $B = 1$\,T.
    } \label{fig:11.5nm_Rxx, Rxy}
\end{figure}

In the inset of Fig.~\ref{fig:11.5nm_Rxx, Rxy}~(a) one could see
the density dependence of the electron mobility $\mu$. The maximum
value of mobility is found to be of about $2.8 \cdot
10^5$~cm$^2$/V$\cdot$s that is significantly higher compare to
both QWs of critical thickness \cite{Dobretsova2016a} as well as
thicker ($d_c = 8\ldots8.3$\,nm) QWs \cite{Gusev2013,
Dantscher2017}, known as 2D TIs. The later inspired us to check if
the QW under investigation also demonstrate topological
properties. The energy gap appears when QW's thickness is smaller
or bigger then critical. The both options are possible in our
case, though only the second one results in the formation of
topological edge channels.

\begin{figure}
    \includegraphics[width=1\columnwidth]{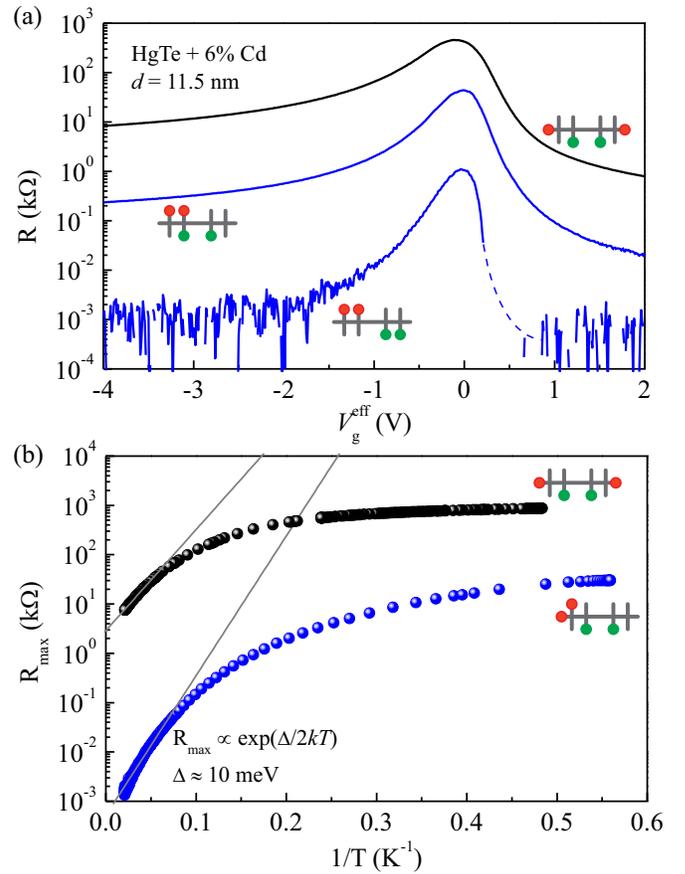}
    \caption{Type 2 QW.
        (a) - The gate voltage dependence of resistance measured at $T = 4.2$~K
        in local geometry (black) and non-local geometries (blue).
        Schematic Hall-bars with source-ground (red circles) and
        probe (green circles) contacts are presented for each geometry.
        (b) - The temperature dependence of local (black circles) and non-local (blue circles)
        resistance's maximum.
        Black solid lines correspond to the fit by the relation $\exp$($\Delta / kT$)
        with $\Delta$ as a fitting parameter.
        Schematic Hall-bars with source-ground (red circles)
        and probe (green circles) contacts are presented for each geometry.
    } \label{fig:11.5_R(T)}
\end{figure}

The non-local transport response is indefeasible sign of the edge
transport \cite{Roth2009, Olshanetsky2015, Olshanetsky2016}. In
order to check the existence of topological states we have
performed a comparison of transport response in local and
non-local geometries. Three gate dependencies of 4-terminal
resistance $R_\text{ij,kl}(V_\text{g}^\text{eff})$ measured in
different geometries are shown in Fig.~\ref{fig:11.5_R(T)}~(a).
Each measured trace is supported by the pictogram explaining the
current and voltage probes configuration. The resistance traces
both in local (black) and non-local (blue) geometries shows
similar behavior with the maximum located in the energy gap. While
the value of the non-local resistance in its maximum is from 1 to
3 orders smaller than the local one, it is still much bigger then
expected in a case of trivial band structure and pure bulk
conductivity. Therefore the non-local resistance results from edge
transport and the QW under investigation is a 2D TI. The absolute
value of non-local resistance in its maximum indicates that the
transport is diffusive (non-ballistic) \cite{Olshanetsky2015,
Olshanetsky2016}.

The most important parameter characterizing 2D TI's band structure
is the energy gap $E_\text{g}$. One could estimate $E_\text{g}$ by
analyzing the temperature dependence of local or non-local
resistance in the gap \cite{Gusev2014, Olshanetsky2015}. By
increasing temperature one increases the number of bulk carriers
that exponentially shunt the measured signal. By fitting the
resistance maximum $R_\text{max}$ with the relation $R_\text{max}
\varpropto$ $\exp$($\Delta / kT$), where $k$ is the Boltzmann
constant and $T$ is the temperature, we have found that the gap is
about 10\,meV (8\,meV from fitting of local resistance and 11\,meV
from non-local one). The obtained value of the gap is
significantly bigger then observed in wide QW-based 2D TIs
\cite{Olshanetsky2015}, making the QW under investigation a
promising candidate for further studies and applications of 2D
TIs.

\section{Conclusion}

In conclusion, we have performed a transport and capacitance
spectroscopy study of two kinds of quantum wells, namely
Cd$_{0.02}$Hg$_{0.98}$Te and Cd$_{0.06}$Hg$_{0.94}$Te with the
thicknesses of 7.4 and 11.5\,nm, accordingly. We have established
that the first quantum well exhibits a massless Dirac fermions
system with a quality slightly better then in conventional HgTe
quantum wells of critical thickness. Second quantum well exhibits
a high-quality two-dimensional topological insulator state with
the energy gap of around 10\,meV and well-defined edge transport
making it as a good candidate for further study and applications
of two-dimensional topological insulators.

\begin{acknowledgments}

The work was supported by RFBR grant No.18-52-16007.

\end{acknowledgments}

\bibliography{library}

\begin{thebibliography}{25}%
\makeatletter
\providecommand \@ifxundefined [1]{%
 \@ifx{#1\undefined}
}%
\providecommand \@ifnum [1]{%
 \ifnum #1\expandafter \@firstoftwo
 \else \expandafter \@secondoftwo
 \fi
}%
\providecommand \@ifx [1]{%
 \ifx #1\expandafter \@firstoftwo
 \else \expandafter \@secondoftwo
 \fi
}%
\providecommand \natexlab [1]{#1}%
\providecommand \enquote  [1]{``#1''}%
\providecommand \bibnamefont  [1]{#1}%
\providecommand \bibfnamefont [1]{#1}%
\providecommand \citenamefont [1]{#1}%
\providecommand \href@noop [0]{\@secondoftwo}%
\providecommand \href [0]{\begingroup \@sanitize@url \@href}%
\providecommand \@href[1]{\@@startlink{#1}\@@href}%
\providecommand \@@href[1]{\endgroup#1\@@endlink}%
\providecommand \@sanitize@url [0]{\catcode `\\12\catcode `\$12\catcode
  `\&12\catcode `\#12\catcode `\^12\catcode `\_12\catcode `\%12\relax}%
\providecommand \@@startlink[1]{}%
\providecommand \@@endlink[0]{}%
\providecommand \url  [0]{\begingroup\@sanitize@url \@url }%
\providecommand \@url [1]{\endgroup\@href {#1}{\urlprefix }}%
\providecommand \urlprefix  [0]{URL }%
\providecommand \Eprint [0]{\href }%
\providecommand \doibase [0]{http://dx.doi.org/}%
\providecommand \selectlanguage [0]{\@gobble}%
\providecommand \bibinfo  [0]{\@secondoftwo}%
\providecommand \bibfield  [0]{\@secondoftwo}%
\providecommand \translation [1]{[#1]}%
\providecommand \BibitemOpen [0]{}%
\providecommand \bibitemStop [0]{}%
\providecommand \bibitemNoStop [0]{.\EOS\space}%
\providecommand \EOS [0]{\spacefactor3000\relax}%
\providecommand \BibitemShut  [1]{\csname bibitem#1\endcsname}%
\let\auto@bib@innerbib\@empty
\bibitem [{\citenamefont {Raichev}(2012)}]{Raichev2012}%
  \BibitemOpen
  \bibfield  {author} {\bibinfo {author} {\bibfnamefont {O.~E.}\ \bibnamefont
  {Raichev}},\ }\bibfield  {title} {\enquote {\bibinfo {title} {{Effective
  Hamiltonian, energy spectrum, and phase transition induced by in-plane
  magnetic field in symmetric HgTe quantum wells}},}\ }\href {\doibase
  10.1103/PhysRevB.85.045310} {\bibfield  {journal} {\bibinfo  {journal} {Phys.
  Rev. B}\ }\textbf {\bibinfo {volume} {85}},\ \bibinfo {pages} {045310}
  (\bibinfo {year} {2012})}\BibitemShut {NoStop}%
\bibitem [{\citenamefont {Buttner}\ \emph {et~al.}(2011)\citenamefont
  {Buttner}, \citenamefont {Liu}, \citenamefont {Tkachov}, \citenamefont
  {Novik}, \citenamefont {Brune}, \citenamefont {Buhmann}, \citenamefont
  {Hankiewicz}, \citenamefont {Recher}, \citenamefont {Trauzettel},
  \citenamefont {Zhang},\ and\ \citenamefont {Molenkamp}}]{Buttner2011}%
  \BibitemOpen
  \bibfield  {author} {\bibinfo {author} {\bibfnamefont {B.}~\bibnamefont
  {Buttner}}, \bibinfo {author} {\bibfnamefont {C.~X.}\ \bibnamefont {Liu}},
  \bibinfo {author} {\bibfnamefont {G.}~\bibnamefont {Tkachov}}, \bibinfo
  {author} {\bibfnamefont {E.~G.}\ \bibnamefont {Novik}}, \bibinfo {author}
  {\bibfnamefont {C.}~\bibnamefont {Brune}}, \bibinfo {author} {\bibfnamefont
  {H.}~\bibnamefont {Buhmann}}, \bibinfo {author} {\bibfnamefont {E.~M.}\
  \bibnamefont {Hankiewicz}}, \bibinfo {author} {\bibfnamefont
  {P.}~\bibnamefont {Recher}}, \bibinfo {author} {\bibfnamefont
  {B.}~\bibnamefont {Trauzettel}}, \bibinfo {author} {\bibfnamefont {S.~C.}\
  \bibnamefont {Zhang}}, \ and\ \bibinfo {author} {\bibfnamefont {L.~W.}\
  \bibnamefont {Molenkamp}},\ }\bibfield  {title} {\enquote {\bibinfo {title}
  {{Single valley Dirac fermions in zero-gap HgTe quantum wells}},}\ }\href
  {\doibase 10.1038/nphys1914} {\bibfield  {journal} {\bibinfo  {journal} {Nat.
  Phys.}\ }\textbf {\bibinfo {volume} {7}},\ \bibinfo {pages} {418} (\bibinfo
  {year} {2011})}\BibitemShut {NoStop}%
\bibitem [{\citenamefont {Tkachov}\ \emph {et~al.}(2011)\citenamefont
  {Tkachov}, \citenamefont {Thienel}, \citenamefont {Pinneker}, \citenamefont
  {B{\"{u}}ttner}, \citenamefont {Br{\"{u}}ne}, \citenamefont {Buhmann},
  \citenamefont {Molenkamp},\ and\ \citenamefont {Hankiewicz}}]{Tkachov2011a}%
  \BibitemOpen
  \bibfield  {author} {\bibinfo {author} {\bibfnamefont {G.}~\bibnamefont
  {Tkachov}}, \bibinfo {author} {\bibfnamefont {C.}~\bibnamefont {Thienel}},
  \bibinfo {author} {\bibfnamefont {V.}~\bibnamefont {Pinneker}}, \bibinfo
  {author} {\bibfnamefont {B.}~\bibnamefont {B{\"{u}}ttner}}, \bibinfo {author}
  {\bibfnamefont {C.}~\bibnamefont {Br{\"{u}}ne}}, \bibinfo {author}
  {\bibfnamefont {H.}~\bibnamefont {Buhmann}}, \bibinfo {author} {\bibfnamefont
  {L.~W.}\ \bibnamefont {Molenkamp}}, \ and\ \bibinfo {author} {\bibfnamefont
  {E.~M.}\ \bibnamefont {Hankiewicz}},\ }\bibfield  {title} {\enquote {\bibinfo
  {title} {{Backscattering of Dirac fermions in HgTe quantum wells with a
  finite gap}},}\ }\href {\doibase 10.1103/PhysRevLett.106.076802} {\bibfield
  {journal} {\bibinfo  {journal} {Phys. Rev. Lett.}\ }\textbf {\bibinfo
  {volume} {106}},\ \bibinfo {pages} {076802} (\bibinfo {year} {2011})},\
  \Eprint {http://arxiv.org/abs/1101.5692} {arXiv:1101.5692} \BibitemShut
  {NoStop}%
\bibitem [{\citenamefont {Kozlov}\ \emph {et~al.}(2013)\citenamefont {Kozlov},
  \citenamefont {Kvon}, \citenamefont {Mikhailov},\ and\ \citenamefont
  {Dvoretsky}}]{Kozlov2012}%
  \BibitemOpen
  \bibfield  {author} {\bibinfo {author} {\bibfnamefont {D.~A.}\ \bibnamefont
  {Kozlov}}, \bibinfo {author} {\bibfnamefont {Z.~D.}\ \bibnamefont {Kvon}},
  \bibinfo {author} {\bibfnamefont {N.~N.}\ \bibnamefont {Mikhailov}}, \ and\
  \bibinfo {author} {\bibfnamefont {S.~A.}\ \bibnamefont {Dvoretsky}},\
  }\bibfield  {title} {\enquote {\bibinfo {title} {{Weak localization of Dirac
  fermions in HgTe quantum wells}},}\ }\href {\doibase
  10.1134/S0021364012230099} {\bibfield  {journal} {\bibinfo  {journal} {JETP
  Lett.}\ }\textbf {\bibinfo {volume} {96}},\ \bibinfo {pages} {730--734}
  (\bibinfo {year} {2013})}\BibitemShut {NoStop}%
\bibitem [{\citenamefont {Dobretsova}\ \emph {et~al.}(2016)\citenamefont
  {Dobretsova}, \citenamefont {Kvon}, \citenamefont {Braginskii}, \citenamefont
  {Entin},\ and\ \citenamefont {Mikhailov}}]{Dobretsova2016a}%
  \BibitemOpen
  \bibfield  {author} {\bibinfo {author} {\bibfnamefont {A.~A.}\ \bibnamefont
  {Dobretsova}}, \bibinfo {author} {\bibfnamefont {Z.~D.}\ \bibnamefont
  {Kvon}}, \bibinfo {author} {\bibfnamefont {L.~S.}\ \bibnamefont
  {Braginskii}}, \bibinfo {author} {\bibfnamefont {M.~V.}\ \bibnamefont
  {Entin}}, \ and\ \bibinfo {author} {\bibfnamefont {N.~N.}\ \bibnamefont
  {Mikhailov}},\ }\bibfield  {title} {\enquote {\bibinfo {title} {{Mobility of
  Dirac electrons in HgTe quantum wells}},}\ }\href {\doibase
  10.7868/S0370274X16180053} {\bibfield  {journal} {\bibinfo  {journal} {JETP
  Lett.}\ }\textbf {\bibinfo {volume} {104}},\ \bibinfo {pages} {388} (\bibinfo
  {year} {2016})}\BibitemShut {NoStop}%
\bibitem [{\citenamefont {Kozlov}\ \emph
  {et~al.}(2014{\natexlab{a}})\citenamefont {Kozlov}, \citenamefont {Kvon},
  \citenamefont {Mikhailov}, \citenamefont {Dvoretsky}, \citenamefont
  {Weish{\"{a}}upl}, \citenamefont {Krupko},\ and\ \citenamefont
  {Portal}}]{Kozlov2014a}%
  \BibitemOpen
  \bibfield  {author} {\bibinfo {author} {\bibfnamefont {D.~A.}\ \bibnamefont
  {Kozlov}}, \bibinfo {author} {\bibfnamefont {Z.~D.}\ \bibnamefont {Kvon}},
  \bibinfo {author} {\bibfnamefont {N.~N.}\ \bibnamefont {Mikhailov}}, \bibinfo
  {author} {\bibfnamefont {S.~A.}\ \bibnamefont {Dvoretsky}}, \bibinfo {author}
  {\bibfnamefont {S.}~\bibnamefont {Weish{\"{a}}upl}}, \bibinfo {author}
  {\bibfnamefont {Y.}~\bibnamefont {Krupko}}, \ and\ \bibinfo {author}
  {\bibfnamefont {J.-C.}\ \bibnamefont {Portal}},\ }\bibfield  {title}
  {\enquote {\bibinfo {title} {{Quantum Hall effect in HgTe quantum wells at
  nitrogen temperatures}},}\ }\href {\doibase 10.1063/1.4896682} {\bibfield
  {journal} {\bibinfo  {journal} {Appl. Phys. Lett.}\ }\textbf {\bibinfo
  {volume} {105}},\ \bibinfo {pages} {132102} (\bibinfo {year}
  {2014}{\natexlab{a}})}\BibitemShut {NoStop}%
\bibitem [{\citenamefont {Kozlov}\ \emph
  {et~al.}(2014{\natexlab{b}})\citenamefont {Kozlov}, \citenamefont {Kvon},
  \citenamefont {Mikhailov},\ and\ \citenamefont {Dvoretsky}}]{Kozlov2014b}%
  \BibitemOpen
  \bibfield  {author} {\bibinfo {author} {\bibfnamefont {D.~A.}\ \bibnamefont
  {Kozlov}}, \bibinfo {author} {\bibfnamefont {Z.~D.}\ \bibnamefont {Kvon}},
  \bibinfo {author} {\bibfnamefont {N.~N.}\ \bibnamefont {Mikhailov}}, \ and\
  \bibinfo {author} {\bibfnamefont {S.~A.}\ \bibnamefont {Dvoretsky}},\
  }\bibfield  {title} {\enquote {\bibinfo {title} {{Quantum Hall Effect in a
  System of Gapless Dirac Fermions in HgTe Quantum Wells}},}\ }\href {\doibase
  https://doi.org/10.1134/S0021364014230076} {\bibfield  {journal} {\bibinfo
  {journal} {JETP Lett.}\ }\textbf {\bibinfo {volume} {100}},\ \bibinfo {pages}
  {724} (\bibinfo {year} {2014}{\natexlab{b}})}\BibitemShut {NoStop}%
\bibitem [{\citenamefont {Kvon}\ \emph {et~al.}(2012)\citenamefont {Kvon},
  \citenamefont {Danilov}, \citenamefont {Kozlov}, \citenamefont {Zoth},
  \citenamefont {Mikhailov}, \citenamefont {Dvoretskii},\ and\ \citenamefont
  {Ganichev}}]{Kvon2011}%
  \BibitemOpen
  \bibfield  {author} {\bibinfo {author} {\bibfnamefont {Z.~D.}\ \bibnamefont
  {Kvon}}, \bibinfo {author} {\bibfnamefont {S.~N.}\ \bibnamefont {Danilov}},
  \bibinfo {author} {\bibfnamefont {D.~A.}\ \bibnamefont {Kozlov}}, \bibinfo
  {author} {\bibfnamefont {C.}~\bibnamefont {Zoth}}, \bibinfo {author}
  {\bibfnamefont {N.~N.}\ \bibnamefont {Mikhailov}}, \bibinfo {author}
  {\bibfnamefont {S.~A.}\ \bibnamefont {Dvoretskii}}, \ and\ \bibinfo {author}
  {\bibfnamefont {S.~D.}\ \bibnamefont {Ganichev}},\ }\bibfield  {title}
  {\enquote {\bibinfo {title} {{Cyclotron Resonance of Dirac Ferions in HgTe
  Quantum Wells}},}\ }\href {\doibase 10.1134/S002136401123007X} {\bibfield
  {journal} {\bibinfo  {journal} {JETP Lett.}\ }\textbf {\bibinfo {volume}
  {94}},\ \bibinfo {pages} {816--819} (\bibinfo {year} {2012})}\BibitemShut
  {NoStop}%
\bibitem [{\citenamefont {Olbrich}\ \emph {et~al.}(2013)\citenamefont
  {Olbrich}, \citenamefont {Zoth}, \citenamefont {Vierling}, \citenamefont
  {Dantscher}, \citenamefont {Budkin}, \citenamefont {Tarasenko}, \citenamefont
  {Bel'kov}, \citenamefont {Kozlov}, \citenamefont {Kvon}, \citenamefont
  {Mikhailov}, \citenamefont {Dvoretsky},\ and\ \citenamefont
  {Ganichev}}]{Olbrich2013a}%
  \BibitemOpen
  \bibfield  {author} {\bibinfo {author} {\bibfnamefont {P.}~\bibnamefont
  {Olbrich}}, \bibinfo {author} {\bibfnamefont {C.}~\bibnamefont {Zoth}},
  \bibinfo {author} {\bibfnamefont {P.}~\bibnamefont {Vierling}}, \bibinfo
  {author} {\bibfnamefont {K.-M.}\ \bibnamefont {Dantscher}}, \bibinfo {author}
  {\bibfnamefont {G.~V.}\ \bibnamefont {Budkin}}, \bibinfo {author}
  {\bibfnamefont {S.~A.}\ \bibnamefont {Tarasenko}}, \bibinfo {author}
  {\bibfnamefont {V.~V.}\ \bibnamefont {Bel'kov}}, \bibinfo {author}
  {\bibfnamefont {D.~A.}\ \bibnamefont {Kozlov}}, \bibinfo {author}
  {\bibfnamefont {Z.~D.}\ \bibnamefont {Kvon}}, \bibinfo {author}
  {\bibfnamefont {N.~N.}\ \bibnamefont {Mikhailov}}, \bibinfo {author}
  {\bibfnamefont {S.~A.}\ \bibnamefont {Dvoretsky}}, \ and\ \bibinfo {author}
  {\bibfnamefont {S.~D.}\ \bibnamefont {Ganichev}},\ }\bibfield  {title}
  {\enquote {\bibinfo {title} {{Giant photocurrents in a Dirac fermion system
  at cyclotron resonance}},}\ }\href {\doibase 10.1103/PhysRevB.87.235439}
  {\bibfield  {journal} {\bibinfo  {journal} {Phys. Rev. B}\ }\textbf {\bibinfo
  {volume} {87}},\ \bibinfo {pages} {235439} (\bibinfo {year} {2013})},\
  \Eprint {http://arxiv.org/abs/1301.4572} {arXiv:1301.4572} \BibitemShut
  {NoStop}%
\bibitem [{\citenamefont {Zoth}\ \emph {et~al.}(2014)\citenamefont {Zoth},
  \citenamefont {Olbrich}, \citenamefont {Vierling}, \citenamefont {Dantscher},
  \citenamefont {Bel'kov}, \citenamefont {Semina}, \citenamefont {Glazov},
  \citenamefont {Golub}, \citenamefont {Kozlov}, \citenamefont {Kvon},
  \citenamefont {Mikhailov}, \citenamefont {Dvoretsky},\ and\ \citenamefont
  {Ganichev}}]{Zoth2014}%
  \BibitemOpen
  \bibfield  {author} {\bibinfo {author} {\bibfnamefont {C.}~\bibnamefont
  {Zoth}}, \bibinfo {author} {\bibfnamefont {P.}~\bibnamefont {Olbrich}},
  \bibinfo {author} {\bibfnamefont {P.}~\bibnamefont {Vierling}}, \bibinfo
  {author} {\bibfnamefont {K.-M.}\ \bibnamefont {Dantscher}}, \bibinfo {author}
  {\bibfnamefont {V.~V.}\ \bibnamefont {Bel'kov}}, \bibinfo {author}
  {\bibfnamefont {M.~A.}\ \bibnamefont {Semina}}, \bibinfo {author}
  {\bibfnamefont {M.~M.}\ \bibnamefont {Glazov}}, \bibinfo {author}
  {\bibfnamefont {L.~E.}\ \bibnamefont {Golub}}, \bibinfo {author}
  {\bibfnamefont {D.~a.}\ \bibnamefont {Kozlov}}, \bibinfo {author}
  {\bibfnamefont {Z.~D.}\ \bibnamefont {Kvon}}, \bibinfo {author}
  {\bibfnamefont {N.~N.}\ \bibnamefont {Mikhailov}}, \bibinfo {author}
  {\bibfnamefont {S.~A.}\ \bibnamefont {Dvoretsky}}, \ and\ \bibinfo {author}
  {\bibfnamefont {S.~D.}\ \bibnamefont {Ganichev}},\ }\bibfield  {title}
  {\enquote {\bibinfo {title} {{Quantum oscillations of photocurrents in HgTe
  quantum wells with Dirac and parabolic dispersions}},}\ }\href {\doibase
  10.1103/PhysRevB.90.205415} {\bibfield  {journal} {\bibinfo  {journal} {Phys.
  Rev. B}\ }\textbf {\bibinfo {volume} {90}},\ \bibinfo {pages} {205415}
  (\bibinfo {year} {2014})}\BibitemShut {NoStop}%
\bibitem [{\citenamefont {Kozlov}\ \emph {et~al.}(2016)\citenamefont {Kozlov},
  \citenamefont {Savchenko}, \citenamefont {Ziegler}, \citenamefont {Kvon},
  \citenamefont {Mikhailov}, \citenamefont {Dvoretsky},\ and\ \citenamefont
  {Weiss}}]{Kozlov2016c}%
  \BibitemOpen
  \bibfield  {author} {\bibinfo {author} {\bibfnamefont {D.~A.}\ \bibnamefont
  {Kozlov}}, \bibinfo {author} {\bibfnamefont {M.~L.}\ \bibnamefont
  {Savchenko}}, \bibinfo {author} {\bibfnamefont {J.}~\bibnamefont {Ziegler}},
  \bibinfo {author} {\bibfnamefont {Z.~D.}\ \bibnamefont {Kvon}}, \bibinfo
  {author} {\bibfnamefont {N.~N.}\ \bibnamefont {Mikhailov}}, \bibinfo {author}
  {\bibfnamefont {S.~A.}\ \bibnamefont {Dvoretsky}}, \ and\ \bibinfo {author}
  {\bibfnamefont {D.}~\bibnamefont {Weiss}},\ }\bibfield  {title} {\enquote
  {\bibinfo {title} {{Capacitance Spectroscopy of a System of Gapless Dirac
  Fermions in a HgTe Quantum Well}},}\ }\href {\doibase
  10.7868/S0370274X16240085} {\bibfield  {journal} {\bibinfo  {journal} {JETP
  Lett.}\ }\textbf {\bibinfo {volume} {104}},\ \bibinfo {pages} {865} (\bibinfo
  {year} {2016})}\BibitemShut {NoStop}%
\bibitem [{\citenamefont {Shuvaev}\ \emph {et~al.}(2016)\citenamefont
  {Shuvaev}, \citenamefont {Dziom}, \citenamefont {Kvon}, \citenamefont
  {Mikhailov},\ and\ \citenamefont {Pimenov}}]{Shuvaev2016}%
  \BibitemOpen
  \bibfield  {author} {\bibinfo {author} {\bibfnamefont {A.}~\bibnamefont
  {Shuvaev}}, \bibinfo {author} {\bibfnamefont {V.}~\bibnamefont {Dziom}},
  \bibinfo {author} {\bibfnamefont {Z.~D.}\ \bibnamefont {Kvon}}, \bibinfo
  {author} {\bibfnamefont {N.~N.}\ \bibnamefont {Mikhailov}}, \ and\ \bibinfo
  {author} {\bibfnamefont {A.}~\bibnamefont {Pimenov}},\ }\bibfield  {title}
  {\enquote {\bibinfo {title} {{Universal Faraday Rotation in HgTe Wells with
  Critical Thickness}},}\ }\href {\doibase 10.1103/PhysRevLett.117.117401}
  {\bibfield  {journal} {\bibinfo  {journal} {Phys. Rev. Lett.}\ }\textbf
  {\bibinfo {volume} {117}},\ \bibinfo {pages} {117401} (\bibinfo {year}
  {2016})},\ \Eprint {http://arxiv.org/abs/1703.05646} {arXiv:1703.05646}
  \BibitemShut {NoStop}%
\bibitem [{\citenamefont {Gusev}\ \emph {et~al.}(2017)\citenamefont {Gusev},
  \citenamefont {Kozlov}, \citenamefont {Levin}, \citenamefont {Kvon},
  \citenamefont {Mikhailov},\ and\ \citenamefont {Dvoretsky}}]{Gusev2017}%
  \BibitemOpen
  \bibfield  {author} {\bibinfo {author} {\bibfnamefont {G.~M.}\ \bibnamefont
  {Gusev}}, \bibinfo {author} {\bibfnamefont {D.~A.}\ \bibnamefont {Kozlov}},
  \bibinfo {author} {\bibfnamefont {A.~D.}\ \bibnamefont {Levin}}, \bibinfo
  {author} {\bibfnamefont {Z.~D.}\ \bibnamefont {Kvon}}, \bibinfo {author}
  {\bibfnamefont {N.~N.}\ \bibnamefont {Mikhailov}}, \ and\ \bibinfo {author}
  {\bibfnamefont {S.~A.}\ \bibnamefont {Dvoretsky}},\ }\bibfield  {title}
  {\enquote {\bibinfo {title} {{Robust helical edge transport at $\nu$=0
  quantum Hall state}},}\ }\href {\doibase 10.1103/PhysRevB.96.045304}
  {\bibfield  {journal} {\bibinfo  {journal} {Phys. Rev. B}\ }\textbf {\bibinfo
  {volume} {96}},\ \bibinfo {pages} {045304} (\bibinfo {year}
  {2017})}\BibitemShut {NoStop}%
\bibitem [{\citenamefont {Hubmann}\ \emph {et~al.}(2018)\citenamefont
  {Hubmann}, \citenamefont {Budkin}, \citenamefont {Dmitriev}, \citenamefont
  {Gebert}, \citenamefont {Belkov}, \citenamefont {Ivchenko}, \citenamefont
  {Baumann}, \citenamefont {Otteneder}, \citenamefont {Kozlov}, \citenamefont
  {Mikhailov}, \citenamefont {Dvoretsky}, \citenamefont {Kvon},\ and\
  \citenamefont {Ganichev}}]{Hubmann2018}%
  \BibitemOpen
  \bibfield  {author} {\bibinfo {author} {\bibfnamefont {S.}~\bibnamefont
  {Hubmann}}, \bibinfo {author} {\bibfnamefont {G.~V.}\ \bibnamefont {Budkin}},
  \bibinfo {author} {\bibfnamefont {A.~P.}\ \bibnamefont {Dmitriev}}, \bibinfo
  {author} {\bibfnamefont {S.}~\bibnamefont {Gebert}}, \bibinfo {author}
  {\bibfnamefont {V.~V.}\ \bibnamefont {Belkov}}, \bibinfo {author}
  {\bibfnamefont {E.~L.}\ \bibnamefont {Ivchenko}}, \bibinfo {author}
  {\bibfnamefont {S.}~\bibnamefont {Baumann}}, \bibinfo {author} {\bibfnamefont
  {M.}~\bibnamefont {Otteneder}}, \bibinfo {author} {\bibfnamefont {D.~A.}\
  \bibnamefont {Kozlov}}, \bibinfo {author} {\bibfnamefont {N.~N.}\
  \bibnamefont {Mikhailov}}, \bibinfo {author} {\bibfnamefont {S.~A.}\
  \bibnamefont {Dvoretsky}}, \bibinfo {author} {\bibfnamefont {Z.~D.}\
  \bibnamefont {Kvon}}, \ and\ \bibinfo {author} {\bibfnamefont {S.~D.}\
  \bibnamefont {Ganichev}},\ }\bibfield  {title} {\enquote {\bibinfo {title}
  {{High frequency impact ionization and nonlinearity of photocurrent induced
  by intense terahertz radiation in HgTe-based quantum well structures}},}\
  }\href {http://arxiv.org/abs/1812.01304} {\ ,\ \bibinfo {pages} {1--12}
  (\bibinfo {year} {2018})},\ \Eprint {http://arxiv.org/abs/1812.01304}
  {arXiv:1812.01304} \BibitemShut {NoStop}%
\bibitem [{\citenamefont {Dobretsova}\ \emph {et~al.}(2015)\citenamefont
  {Dobretsova}, \citenamefont {Braginskii}, \citenamefont {Entin},
  \citenamefont {Kvon}, \citenamefont {Mikhailov},\ and\ \citenamefont
  {Dvoretsky}}]{Dobretsova2015a}%
  \BibitemOpen
  \bibfield  {author} {\bibinfo {author} {\bibfnamefont {A.~A.}\ \bibnamefont
  {Dobretsova}}, \bibinfo {author} {\bibfnamefont {L.~S.}\ \bibnamefont
  {Braginskii}}, \bibinfo {author} {\bibfnamefont {M.~V.}\ \bibnamefont
  {Entin}}, \bibinfo {author} {\bibfnamefont {Z.~D.}\ \bibnamefont {Kvon}},
  \bibinfo {author} {\bibfnamefont {N.~N.}\ \bibnamefont {Mikhailov}}, \ and\
  \bibinfo {author} {\bibfnamefont {S.~A.}\ \bibnamefont {Dvoretsky}},\
  }\bibfield  {title} {\enquote {\bibinfo {title} {{Surface states in a HgTe
  quantum well and scattering by surface roughness}},}\ }\href {\doibase
  10.7868/S0370274X15050094} {\bibfield  {journal} {\bibinfo  {journal} {JETP
  Lett.}\ }\textbf {\bibinfo {volume} {101}},\ \bibinfo {pages} {330} (\bibinfo
  {year} {2015})}\BibitemShut {NoStop}%
\bibitem [{\citenamefont {Orlita}\ \emph {et~al.}(2014)\citenamefont {Orlita},
  \citenamefont {Basko}, \citenamefont {Zholudev}, \citenamefont {Teppe},
  \citenamefont {Knap}, \citenamefont {Gavrilenko}, \citenamefont {Mikhailov},
  \citenamefont {Dvoretsky}, \citenamefont {Neugebauer}, \citenamefont
  {Faugeras}, \citenamefont {Barra}, \citenamefont {Martinez},\ and\
  \citenamefont {Potemski}}]{Orlita2014}%
  \BibitemOpen
  \bibfield  {author} {\bibinfo {author} {\bibfnamefont {M.}~\bibnamefont
  {Orlita}}, \bibinfo {author} {\bibfnamefont {D.~M.}\ \bibnamefont {Basko}},
  \bibinfo {author} {\bibfnamefont {M.~S.}\ \bibnamefont {Zholudev}}, \bibinfo
  {author} {\bibfnamefont {F.}~\bibnamefont {Teppe}}, \bibinfo {author}
  {\bibfnamefont {W.}~\bibnamefont {Knap}}, \bibinfo {author} {\bibfnamefont
  {V.~I.}\ \bibnamefont {Gavrilenko}}, \bibinfo {author} {\bibfnamefont
  {N.~N.}\ \bibnamefont {Mikhailov}}, \bibinfo {author} {\bibfnamefont {S.~A.}\
  \bibnamefont {Dvoretsky}}, \bibinfo {author} {\bibfnamefont {P.}~\bibnamefont
  {Neugebauer}}, \bibinfo {author} {\bibfnamefont {C.}~\bibnamefont
  {Faugeras}}, \bibinfo {author} {\bibfnamefont {A.-L.}\ \bibnamefont {Barra}},
  \bibinfo {author} {\bibfnamefont {G.}~\bibnamefont {Martinez}}, \ and\
  \bibinfo {author} {\bibfnamefont {M.}~\bibnamefont {Potemski}},\ }\bibfield
  {title} {\enquote {\bibinfo {title} {{Observation of three-dimensional
  massless Kane fermions in a zinc-blende crystal}},}\ }\href {\doibase
  10.1038/nphys2857} {\bibfield  {journal} {\bibinfo  {journal} {Nat. Phys.}\
  }\textbf {\bibinfo {volume} {10}},\ \bibinfo {pages} {233} (\bibinfo {year}
  {2014})}\BibitemShut {NoStop}%
\bibitem [{\citenamefont {Teppe}\ \emph {et~al.}(2016)\citenamefont {Teppe},
  \citenamefont {Marcinkiewicz}, \citenamefont {Krishtopenko}, \citenamefont
  {Ruffenach}, \citenamefont {Consejo}, \citenamefont {Kadykov}, \citenamefont
  {Desrat}, \citenamefont {But}, \citenamefont {Knap}, \citenamefont {Ludwig},
  \citenamefont {Moon}, \citenamefont {Smirnov}, \citenamefont {Orlita},
  \citenamefont {Jiang}, \citenamefont {Morozov}, \citenamefont {Gavrilenko},
  \citenamefont {Mikhailov},\ and\ \citenamefont {Dvoretskii}}]{Teppe2016}%
  \BibitemOpen
  \bibfield  {author} {\bibinfo {author} {\bibfnamefont {F.}~\bibnamefont
  {Teppe}}, \bibinfo {author} {\bibfnamefont {M.}~\bibnamefont
  {Marcinkiewicz}}, \bibinfo {author} {\bibfnamefont {S.~S.}\ \bibnamefont
  {Krishtopenko}}, \bibinfo {author} {\bibfnamefont {S.}~\bibnamefont
  {Ruffenach}}, \bibinfo {author} {\bibfnamefont {C.}~\bibnamefont {Consejo}},
  \bibinfo {author} {\bibfnamefont {A.~M.}\ \bibnamefont {Kadykov}}, \bibinfo
  {author} {\bibfnamefont {W.}~\bibnamefont {Desrat}}, \bibinfo {author}
  {\bibfnamefont {D.}~\bibnamefont {But}}, \bibinfo {author} {\bibfnamefont
  {W.}~\bibnamefont {Knap}}, \bibinfo {author} {\bibfnamefont {J.}~\bibnamefont
  {Ludwig}}, \bibinfo {author} {\bibfnamefont {S.}~\bibnamefont {Moon}},
  \bibinfo {author} {\bibfnamefont {D.}~\bibnamefont {Smirnov}}, \bibinfo
  {author} {\bibfnamefont {M.}~\bibnamefont {Orlita}}, \bibinfo {author}
  {\bibfnamefont {Z.}~\bibnamefont {Jiang}}, \bibinfo {author} {\bibfnamefont
  {S.~V.}\ \bibnamefont {Morozov}}, \bibinfo {author} {\bibfnamefont
  {V.}~\bibnamefont {Gavrilenko}}, \bibinfo {author} {\bibfnamefont {N.~N.}\
  \bibnamefont {Mikhailov}}, \ and\ \bibinfo {author} {\bibfnamefont {S.~A.}\
  \bibnamefont {Dvoretskii}},\ }\bibfield  {title} {\enquote {\bibinfo {title}
  {{Temperature-driven massless Kane fermions in HgCdTe crystals}},}\ }\href
  {\doibase 10.1038/ncomms12576} {\bibfield  {journal} {\bibinfo  {journal}
  {Nat. Commun.}\ }\textbf {\bibinfo {volume} {7}},\ \bibinfo {pages} {12576}
  (\bibinfo {year} {2016})},\ \Eprint {http://arxiv.org/abs/1602.50999}
  {arXiv:1602.50999} \BibitemShut {NoStop}%
\bibitem [{\citenamefont {Smith}\ \emph {et~al.}(1985)\citenamefont {Smith},
  \citenamefont {Goldberg}, \citenamefont {Stiles},\ and\ \citenamefont
  {Heiblum}}]{Smith1985}%
  \BibitemOpen
  \bibfield  {author} {\bibinfo {author} {\bibfnamefont {T.~P.}\ \bibnamefont
  {Smith}}, \bibinfo {author} {\bibfnamefont {B.~B.}\ \bibnamefont {Goldberg}},
  \bibinfo {author} {\bibfnamefont {P.~J.}\ \bibnamefont {Stiles}}, \ and\
  \bibinfo {author} {\bibfnamefont {M.}~\bibnamefont {Heiblum}},\ }\bibfield
  {title} {\enquote {\bibinfo {title} {{Direct measurement of the density of
  states of a two-dimensional electron gas}},}\ }\href {\doibase
  10.1103/PhysRevB.32.2696} {\bibfield  {journal} {\bibinfo  {journal} {Phys.
  Rev. B}\ }\textbf {\bibinfo {volume} {32}},\ \bibinfo {pages} {2696--2699}
  (\bibinfo {year} {1985})}\BibitemShut {NoStop}%
\bibitem [{\citenamefont {Ponomarenko}\ \emph {et~al.}(2010)\citenamefont
  {Ponomarenko}, \citenamefont {Yang}, \citenamefont {Gorbachev}, \citenamefont
  {Blake}, \citenamefont {Mayorov}, \citenamefont {Novoselov}, \citenamefont
  {Katsnelson},\ and\ \citenamefont {Geim}}]{Ponomarenko2010}%
  \BibitemOpen
  \bibfield  {author} {\bibinfo {author} {\bibfnamefont {L.~A.}\ \bibnamefont
  {Ponomarenko}}, \bibinfo {author} {\bibfnamefont {R.}~\bibnamefont {Yang}},
  \bibinfo {author} {\bibfnamefont {R.~V.}\ \bibnamefont {Gorbachev}}, \bibinfo
  {author} {\bibfnamefont {P.}~\bibnamefont {Blake}}, \bibinfo {author}
  {\bibfnamefont {A.~S.}\ \bibnamefont {Mayorov}}, \bibinfo {author}
  {\bibfnamefont {K.~S.}\ \bibnamefont {Novoselov}}, \bibinfo {author}
  {\bibfnamefont {M.~I.}\ \bibnamefont {Katsnelson}}, \ and\ \bibinfo {author}
  {\bibfnamefont {A.~K.}\ \bibnamefont {Geim}},\ }\bibfield  {title} {\enquote
  {\bibinfo {title} {{Density of States and Zero Landau Level Probed through
  Capacitance of Graphene}},}\ }\href {\doibase 10.1103/PhysRevLett.105.136801}
  {\bibfield  {journal} {\bibinfo  {journal} {Phys. Rev. Lett.}\ }\textbf
  {\bibinfo {volume} {105}},\ \bibinfo {pages} {136801} (\bibinfo {year}
  {2010})},\ \Eprint {http://arxiv.org/abs/1005.4793} {arXiv:1005.4793}
  \BibitemShut {NoStop}%
\bibitem [{\citenamefont {Gusev}\ \emph {et~al.}(2013)\citenamefont {Gusev},
  \citenamefont {Olshanetsky}, \citenamefont {Kvon}, \citenamefont
  {Mikhailov},\ and\ \citenamefont {Dvoretsky}}]{Gusev2013}%
  \BibitemOpen
  \bibfield  {author} {\bibinfo {author} {\bibfnamefont {G.~M.}\ \bibnamefont
  {Gusev}}, \bibinfo {author} {\bibfnamefont {E.~B.}\ \bibnamefont
  {Olshanetsky}}, \bibinfo {author} {\bibfnamefont {Z.~D.}\ \bibnamefont
  {Kvon}}, \bibinfo {author} {\bibfnamefont {N.~N.}\ \bibnamefont {Mikhailov}},
  \ and\ \bibinfo {author} {\bibfnamefont {S.~A.}\ \bibnamefont {Dvoretsky}},\
  }\bibfield  {title} {\enquote {\bibinfo {title} {{Linear magnetoresistance in
  HgTe quantum wells}},}\ }\href {\doibase 10.1103/PhysRevB.87.081311}
  {\bibfield  {journal} {\bibinfo  {journal} {Phys. Rev. B}\ }\textbf {\bibinfo
  {volume} {87}},\ \bibinfo {pages} {081311} (\bibinfo {year} {2013})},\
  \Eprint {http://arxiv.org/abs/1302.6754} {arXiv:1302.6754} \BibitemShut
  {NoStop}%
\bibitem [{\citenamefont {Dantscher}\ \emph {et~al.}(2017)\citenamefont
  {Dantscher}, \citenamefont {Kozlov}, \citenamefont {Scherr}, \citenamefont
  {Gebert}, \citenamefont {B{\"{a}}renf{\"{a}}nger}, \citenamefont {Durnev},
  \citenamefont {Tarasenko}, \citenamefont {Bel'kov}, \citenamefont
  {Mikhailov}, \citenamefont {Dvoretsky}, \citenamefont {Kvon}, \citenamefont
  {Ziegler}, \citenamefont {Weiss},\ and\ \citenamefont
  {Ganichev}}]{Dantscher2017}%
  \BibitemOpen
  \bibfield  {author} {\bibinfo {author} {\bibfnamefont {K.-M.}\ \bibnamefont
  {Dantscher}}, \bibinfo {author} {\bibfnamefont {D.~A.}\ \bibnamefont
  {Kozlov}}, \bibinfo {author} {\bibfnamefont {M.~T.}\ \bibnamefont {Scherr}},
  \bibinfo {author} {\bibfnamefont {S.}~\bibnamefont {Gebert}}, \bibinfo
  {author} {\bibfnamefont {J.}~\bibnamefont {B{\"{a}}renf{\"{a}}nger}},
  \bibinfo {author} {\bibfnamefont {M.~V.}\ \bibnamefont {Durnev}}, \bibinfo
  {author} {\bibfnamefont {S.~A.}\ \bibnamefont {Tarasenko}}, \bibinfo {author}
  {\bibfnamefont {V.~V.}\ \bibnamefont {Bel'kov}}, \bibinfo {author}
  {\bibfnamefont {N.~N.}\ \bibnamefont {Mikhailov}}, \bibinfo {author}
  {\bibfnamefont {S.~A.}\ \bibnamefont {Dvoretsky}}, \bibinfo {author}
  {\bibfnamefont {Z.~D.}\ \bibnamefont {Kvon}}, \bibinfo {author}
  {\bibfnamefont {J.}~\bibnamefont {Ziegler}}, \bibinfo {author} {\bibfnamefont
  {D.}~\bibnamefont {Weiss}}, \ and\ \bibinfo {author} {\bibfnamefont {S.~D.}\
  \bibnamefont {Ganichev}},\ }\bibfield  {title} {\enquote {\bibinfo {title}
  {{Photogalvanic probing of helical edge channels in two-dimensional HgTe
  topological insulators}},}\ }\href {\doibase 10.1103/PhysRevB.95.201103}
  {\bibfield  {journal} {\bibinfo  {journal} {Phys. Rev. B}\ }\textbf {\bibinfo
  {volume} {95}},\ \bibinfo {pages} {201103} (\bibinfo {year} {2017})},\
  \Eprint {http://arxiv.org/abs/1612.08854} {arXiv:1612.08854} \BibitemShut
  {NoStop}%
\bibitem [{\citenamefont {Roth}\ \emph {et~al.}(2009)\citenamefont {Roth},
  \citenamefont {Brune}, \citenamefont {Buhmann}, \citenamefont {Molenkamp},
  \citenamefont {Maciejko}, \citenamefont {Qi},\ and\ \citenamefont
  {Zhang}}]{Roth2009}%
  \BibitemOpen
  \bibfield  {author} {\bibinfo {author} {\bibfnamefont {A.}~\bibnamefont
  {Roth}}, \bibinfo {author} {\bibfnamefont {C.}~\bibnamefont {Brune}},
  \bibinfo {author} {\bibfnamefont {H.}~\bibnamefont {Buhmann}}, \bibinfo
  {author} {\bibfnamefont {L.~W.}\ \bibnamefont {Molenkamp}}, \bibinfo {author}
  {\bibfnamefont {J.}~\bibnamefont {Maciejko}}, \bibinfo {author}
  {\bibfnamefont {X.-L.}\ \bibnamefont {Qi}}, \ and\ \bibinfo {author}
  {\bibfnamefont {S.-C.}\ \bibnamefont {Zhang}},\ }\bibfield  {title} {\enquote
  {\bibinfo {title} {{Nonlocal Transport in the Quantum Spin Hall State}},}\
  }\href {\doibase 10.1126/science.1174736} {\bibfield  {journal} {\bibinfo
  {journal} {Science}\ }\textbf {\bibinfo {volume} {325}},\ \bibinfo {pages}
  {294--297} (\bibinfo {year} {2009})},\ \Eprint
  {http://arxiv.org/abs/0905.0365} {arXiv:0905.0365} \BibitemShut {NoStop}%
\bibitem [{\citenamefont {Olshanetsky}\ \emph {et~al.}(2015)\citenamefont
  {Olshanetsky}, \citenamefont {Kvon}, \citenamefont {Gusev}, \citenamefont
  {Levin}, \citenamefont {Raichev}, \citenamefont {Mikhailov},\ and\
  \citenamefont {Dvoretsky}}]{Olshanetsky2015}%
  \BibitemOpen
  \bibfield  {author} {\bibinfo {author} {\bibfnamefont {E.~B.}\ \bibnamefont
  {Olshanetsky}}, \bibinfo {author} {\bibfnamefont {Z.~D.}\ \bibnamefont
  {Kvon}}, \bibinfo {author} {\bibfnamefont {G.~M.}\ \bibnamefont {Gusev}},
  \bibinfo {author} {\bibfnamefont {A.~D.}\ \bibnamefont {Levin}}, \bibinfo
  {author} {\bibfnamefont {O.~E.}\ \bibnamefont {Raichev}}, \bibinfo {author}
  {\bibfnamefont {N.~N.}\ \bibnamefont {Mikhailov}}, \ and\ \bibinfo {author}
  {\bibfnamefont {S.~A.}\ \bibnamefont {Dvoretsky}},\ }\bibfield  {title}
  {\enquote {\bibinfo {title} {{Persistence of a Two-Dimensional Topological
  Insulator State in Wide HgTe Quantum Wells}},}\ }\href {\doibase
  10.1103/PhysRevLett.114.126802} {\bibfield  {journal} {\bibinfo  {journal}
  {Phys. Rev. Lett.}\ }\textbf {\bibinfo {volume} {114}},\ \bibinfo {pages}
  {126802} (\bibinfo {year} {2015})},\ \Eprint
  {http://arxiv.org/abs/1503.08978} {arXiv:1503.08978} \BibitemShut {NoStop}%
\bibitem [{\citenamefont {Olshanetsky}\ \emph {et~al.}(2016)\citenamefont
  {Olshanetsky}, \citenamefont {Kvon}, \citenamefont {Gusev}, \citenamefont
  {Mikhailov},\ and\ \citenamefont {Dvoretsky}}]{Olshanetsky2016}%
  \BibitemOpen
  \bibfield  {author} {\bibinfo {author} {\bibfnamefont {E.~B.}\ \bibnamefont
  {Olshanetsky}}, \bibinfo {author} {\bibfnamefont {Z.~D.}\ \bibnamefont
  {Kvon}}, \bibinfo {author} {\bibfnamefont {G.~M.}\ \bibnamefont {Gusev}},
  \bibinfo {author} {\bibfnamefont {N.~N.}\ \bibnamefont {Mikhailov}}, \ and\
  \bibinfo {author} {\bibfnamefont {S.~A.}\ \bibnamefont {Dvoretsky}},\
  }\bibfield  {title} {\enquote {\bibinfo {title} {{Low field magnetoresistance
  in a 2D topological insulator based on wide HgTe quantum well}},}\ }\href
  {\doibase 10.1088/0953-8984/28/34/345801} {\bibfield  {journal} {\bibinfo
  {journal} {J. Phys. Condens. Matter}\ }\textbf {\bibinfo {volume} {28}},\
  \bibinfo {pages} {345801} (\bibinfo {year} {2016})}\BibitemShut {NoStop}%
\bibitem [{\citenamefont {Gusev}\ \emph {et~al.}(2014)\citenamefont {Gusev},
  \citenamefont {Kvon}, \citenamefont {Olshanetsky}, \citenamefont {Levin},
  \citenamefont {Krupko}, \citenamefont {Portal}, \citenamefont {Mikhailov},\
  and\ \citenamefont {Dvoretsky}}]{Gusev2014}%
  \BibitemOpen
  \bibfield  {author} {\bibinfo {author} {\bibfnamefont {G.~M.}\ \bibnamefont
  {Gusev}}, \bibinfo {author} {\bibfnamefont {Z.~D.}\ \bibnamefont {Kvon}},
  \bibinfo {author} {\bibfnamefont {E.~B.}\ \bibnamefont {Olshanetsky}},
  \bibinfo {author} {\bibfnamefont {A.~D.}\ \bibnamefont {Levin}}, \bibinfo
  {author} {\bibfnamefont {Y.}~\bibnamefont {Krupko}}, \bibinfo {author}
  {\bibfnamefont {J.~C.}\ \bibnamefont {Portal}}, \bibinfo {author}
  {\bibfnamefont {N.~N.}\ \bibnamefont {Mikhailov}}, \ and\ \bibinfo {author}
  {\bibfnamefont {S.~A.}\ \bibnamefont {Dvoretsky}},\ }\bibfield  {title}
  {\enquote {\bibinfo {title} {{Temperature dependence of the resistance of a
  two-dimensional topological insulator in a HgTe quantum well}},}\ }\href
  {\doibase 10.1103/PhysRevB.89.125305} {\bibfield  {journal} {\bibinfo
  {journal} {Phys. Rev. B}\ }\textbf {\bibinfo {volume} {89}},\ \bibinfo
  {pages} {125305} (\bibinfo {year} {2014})}\BibitemShut {NoStop}%
\end{thebibliography}%

\end{document}